\documentclass{aa}  
\usepackage{lineno} 
\usepackage{amsmath}
\usepackage{graphicx}
\usepackage{adjustbox}
\usepackage{longtable}
\usepackage{geometry} 
\usepackage{array}
\usepackage{caption} 
\usepackage{subcaption} 
\usepackage{lscape} 
\usepackage{hyperref}
\usepackage[utf8]{inputenc}

\usepackage{txfonts}
\begin{document}

\nolinenumbers
\def\simlt{\lower.5ex\hbox{\ltsima}}
\def\simgt{\lower.5ex\hbox{\gtsima}}

\def\farcm{\hbox{$\mkern-4mu^\prime$}}
\def\parcm{${}^{\prime}$\llap{.}}
\def\farcs{\hbox{$^{\prime\prime}$}~}
\def\parcs{${}^{\prime\prime}$\llap{.}}
\def\km{{\rm\,km}}
\def\kms{{\rm\,km\,s^{-1}}}
\def\mas{{\rm\,mas}}
\def\masyr{{\rm\,mas/yr}}
\def\kpc{{\rm\,kpc}}
\def\mpc{{\rm\,Mpc}}
\def\msun{{\rm\,M_\odot}}
\def\lsun{{\rm\,L_\odot}}
\def\rsun{{\rm\,R_\odot}}
\def\pc{{\rm\,pc}}
\def\cm{{\rm\,cm}}
\def\yr{{\rm\,yr}}
\def\au{{\rm\,AU}}
\def\g{{\rm\,g}}
\def\om{\Omega_0}
\def \ca {{\it ca.\/}}
\def \r {r$^{1/4}$ }
\def \magnitude {$^{\rm m}$}
\def\kr{${\cal K}_r$}
\def\kz{${\cal K}_z$}
\def\kzz{${\cal K}_z(z)$}
\def\mss{{\rm M}_\odot \rm pc^{-2}}
\def\msss{{\rm M}_\odot \rm pc^{-3}}
\newcommand{\fmmm}[1]{\mbox{$#1$}}
\newcommand{\scnd}{\mbox{\fmmm{''}\hskip-0.3em .}}
\newcommand{\scnp}{\mbox{\fmmm{''}}}
\newcommand{\mcnd}{\mbox{\fmmm{'}\hskip-0.3em .}}
\def\Aa{\; \buildrel \circ \over {\rm A}}
\def\AA{$\; \buildrel \circ \over {\rm A}$}
\def\yr{{\rm yr}}
\def\tmaxo{$T_{max}^{opt}$} 
\def\tmeano{$T_{mean}^{opt}$} 
\def\tminv{$T_{min}^{RV}$} 
\def\tmeanv{$T_{mean}^{RV}$} 
\def\tminvfe{$T_{min}^{RV(Fe)}$} 
\def\tmeanvfe{$T_{mean}^{RV(Fe)}$} 
\def\tminvhb{$T_{min}^{RV(H\beta)}$} 
\def\tmeanvhb{$T_{mean}^{RV(H\beta)}$} 

\def\gtsim{\;\lower.6ex\hbox{$\sim$}\kern-6.7pt\raise.4ex\hbox{$>$}\;}
\def\ltsim{\;\lower.6ex\hbox{$\sim$}\kern-6.9pt\raise.4ex\hbox{$<$}\;}
\def\hr{${}^{\hbox{\footnotesize h}}$}
\def\tm{${}^{\hbox{\footnotesize m}}$}
\def\ts{${}^{\hbox{\footnotesize s}}$}
\def\Ts{${}^{\hbox{\footnotesize s}}$\llap{.}}
\def\Deg{${}^\circ$\llap{.}}
\def\Min{${}^{\prime}$\llap{.}}
\def\Sec{${}^{\prime\prime}$\llap{.}}
\def\deg{${}^\circ$}
\def\min{${}^{\prime}$}
\def\sec{${}^{\prime\prime}$}
\def\hst{{\it HST\/}}
\def\gaia{{\it Gaia\/}}
\def\bmv{\hbox{\it B--V\/}}
\def\bmr{\hbox{\it B--R\/}}
\def\bmi{\hbox{\it B--I\/}}
\def\vmi{\hbox{\it V--I\/}}
\def\umi{\hbox{\it U--I\/}}
\def\jmk{\hbox{\it J--K\/}}
\def\bmk{\hbox{\it B--K\/}}
\def\\gc#1{\hbox{NGC$\,$#1}}
\def\wcen{$\omega$ Cen~}
\def\trisv{$T_{ris}^{V}$}
\def\trisg{$T_{ris}^{g}$}
\def\trisr{$T_{ris}^{r}$}
\def\trisi{$T_{ris}^{i}$}
\def\trisz{$T_{ris}^{z}$}
\def\tris{$T_{ris}$}

\setlength{\tabcolsep}{3pt}

\newcommand{\rrl}{RR~Lyr\ae}
\newcommand{\GB}[1]{\textcolor{cyan!100}{#1$_{_{\mathbf{GB}}}$}}
\newcommand{\VFB}[1]{\textcolor{red!100}{#1$_{_{\mathbf{VFB}}}$}}

\title{New theoretical predictions on Type II Cepheids: towards a self consistent Pop. II distance scale}


\nolinenumbers

\author{
M.~Marconi \inst{1}
\and R.~Molinaro \inst{1}
\and V. ~Ripepi \inst{1}
\and G. ~De Somma \inst{1, 6, 7}
\and T. ~Sicignano \inst{2,3}
\and M. ~Deka \inst{1}
\and M. ~Di Criscienzo \inst{4}
\and E. ~Luongo \inst{5,1}
\and I. ~Musella \inst{1}
\and E. ~Trentin \inst{1}
}
\institute{
 INAF-Osservatorio Astronomico di Capodimonte, Salita Moiariello 16, 80131 Napoli, Italy
\and European Southern Observatory, Karl-Schwarzschild-Str. 2, 85748 Garching bei München, Germany
\and Scuola Superiore Meridionale, Largo S. Marcellino 10, 80138 Napoli, Italy
\and INAF-Osservatorio Astronomico di Roma, via Frascati 33, I-00078, Monteporzio Catone, Roma, Italy
\and Università di Salerno, Dipartimento di Fisica “E.R. Caianiello”, Via Giovanni Paolo II 132, 84084 Fisciano (SA), Italy
\and INAF-Osservatorio Astronomico d'Abruzzo, Via Maggini sn, 64100 Teramo, Italy
\and Istituto Nazionale di Fisica Nucleare (INFN) - Sez. di Napoli, Compl. Univ.di Monte S. Angelo, Edificio G, Via Cinthia, I-80126, Napoli, Italy
}

\abstract
{Type II Cepheids are pulsating stars that can be used as standard candles for old stellar populations due to their characteristic Period-Luminosity and Period-Luminosity-Colour relations. They are traditionally divided in 3 sub-classes, namely BL Her, W Vir and RV Tauri.}{In this paper we focus on the first two sub-classes, to provide a new theoretical scenario and develop tools and relations to be adopted in distance scale and old stellar populations studies.}{We have built new nonlinear convective pulsation models of Type II Cepheids, computed along selected stellar evolution tracks and spanning a wide range of pulsation period and stellar parameters. Three chemical compositions have been taken into account, namely $Z=0.0001$ $Y=0.245$,  $Z=0.001$ $Y=0.245$ and $Z=0.01$ $Y=0.26$. For each assumed $Z$ and $Y$, models have been computed following stellar evolution predictions for off Zero Age Horizontal Branch evolution of stellar masses lower than typical RR Lyrae stars, crossing the classical instability strip as BL Her or W Vir pulsating stars.}{A new theoretical prediction for the instability strip boundaries of these classes of variable stars has been obtained together with their dependence on metal abundance. The predicted light and radial velocity curves have been computed along the evolution inside the strip, showing how the amplitude and the morphology are affected by the position relative to the edges and by the luminosity and mass values.
The transformation of bolometric light curves into various photometric systems allowed us to provide new theoretical Period-Luminosity and Period-Wesenheit relations for BL Her and W Vir.}{These relations are found to be consistent with previously published RR Lyrae model results but with a smaller metallicity dependence. Moreover, the application of the inferred theoretical relations to Magellanic and Galactic Type II Cepheid data provides results in good agreement with some independent distance estimates in the literature.} 


\maketitle

\section{Introduction} \label{sect_intro}
\nolinenumbers
 One of the most debated topic of nowadays astrophysics is the so called Hubble constant tension, that is the discrepancy at the 4-5 $\sigma$ level between the early Universe determinations of the Hubble constant $H_0$ through Planck satellite measurement of the Cosmic Microwave Background combined with flat Cold Dark Matter (CDM) theory \citep{Planck2020} and the values obtained by using Classical Cepheids as primary and Supernovae Ia type as secondary distance indicators, in the calibration of the cosmic distance ladder \citep[see e.g.][and references therein]{verde19,riess22}.
 From the stellar point of view, in order to clarify the origin of this tension, we need to accurately investigate possible residual systematics affecting the calibration of the cosmic distance scale, also exploring the capabilities of alternative primary distance indicators.
 Recent results based on the adoption of the Tip of the Red Giant Branch (TRGB) to calibrate the subsequent rung of Type Ia supernovae (SNIa), in place of classical Cepheids,  pointed towards a value of $H_0$ somewhat intermediate between early and late universe results, thus significantly reducing the tension \citep[see e.g.][and references therein]{Freedman2021}, even if subsequent reanalysis of the method, e.g. by \citet{Scolnic23} seemed to confirm the value by \citet[][]{riess22}.
 More recently, \citet{Taylor25} claimed that the adoption of JWST measurements of the TRGB to calibrate SNIa in external galaxies essentially removes the tension. 
 
 In the context of investigating Population II standard candles, in \citet{sicignano24} we used new empirical calibrations of Type II Cepheids (T2Cs) Period-Luminosity (PL) and Period-Wesenheit (PW) relations to obtain distances to 22 GGCs, providing strong support for using these pulsating stars together with the TRGB for cosmic distance scale studies \citep[see also][and references therein]{Baade58,Bono16,Bhardwaj2021}.
  T2Cs are brighter by more than 0.2 dex than the extensively studied RR Lyrae (RRL) class and have been identified in a variety of environments: the Galactic Bulge \citep[see e.g.][]{Matsunaga13,Braga19,soszynski11,soszynski17}, the GGCs \citep[GGCs, see][]{matsunaga2006,ngeow22}, the Galactic halo \citep[][]{Ripepi2023}, and the Magellanic Clouds \citep[see e.g.][]{ripepi2015,soszynski18,sicignano24}.  According to their pulsation period, T2Cs can be classified into three sub-groups \citep[see e.g.][]{soszynski08,bono20}: BL Herculis (BL Her) stars with periods longer than RRLs and shorter than five days; W Virginis (WVir) stars with periods between four and twenty days, RV Tauri stars with periods longer than twenty days. 
 From the theoretical point of view, stellar evolution models for T2Cs were first produced by \citet{Gingold76,Gingold85}, spanning a broad range of stellar masses and chemical compositions. More recently, BL Her stellar evolution and pulsation properties have been investigated by \citet{bono1997nonlinear,dicriscienzo07,marconi07,bono20}.
Here we plan to extend the pulsation analysis computing for the first time the oscillation properties of post Zero Age Horizontal Branch (ZAHB) stars along their evolutionary tracks,
for three different metallicity values, following the detailed theoretical framework provided by \citet{bono20}. Here we exclude W Vir with periods longer than about 10 days and RV Tauri stars. In particular, for the latter it is difficult to constrain their upper mass limits (see Bódi and Kiss 2019), and their evolutionary origins appear highly diverse—spanning both young massive stars and evolved binaries (Manick et al. 2018). 
The paper is structured as follows: in Section~\ref{sect_models}, we present the new set of pulsation models. The predicted instability strip for both BL Her and W Vir pulsating stars as a function of the assumed metal abundance, the new pulsation relations and the predicted atlas of light and radial velocity curves are presented in Section~\ref{sect_results}. 
In Section~\ref{sect_modeled_relations} we derive new theoretical metal-dependent Period-Luminosity-Mass-Temperature, Period-Luminosity and Period-Wesenheit relations, whereas in Section~\ref{sect_observed_relations} we discuss the application to samples of Magellanic and Galactic pulsators. Finally, we report our conclusions in Section~\ref{sect_conclusions}.

\begin{table}[t]
\centering
\caption{The structural parameters of all computed models.} 
\label{table:allModels}
\begin{tabular}{cccccc}
  \hline
Z& Y & M/$M_{\odot}$ & logL/$L_{\odot}$ & T$_{eff}$ & P \\ 
  \hline
0.0001 & 0.245 & 0.530 & 2.450 & 6350 & 3.4978 \\ 
  0.0001 & 0.245 & 0.530 & 2.460 & 6200 & 3.9480 \\ 
  0.0001 & 0.245 & 0.530 & 2.460 & 6100 & 4.2533 \\ 
      ..........\\
  0.0100 & 0.260 & 0.550 & 1.780 & 5682 & 1.3159 \\ 
  0.0100 & 0.260 & 0.550 & 1.790 & 5607 & 1.4047 \\ 
  0.0100 & 0.260 & 0.550 & 1.830 & 5432 & 1.6689 \\ 
   \hline   
\end{tabular}
 \tablefoot{Metallicity (Z); Helium abundance (Y); Mass in solar unit (M/$M_{\odot}$); Luminosity in solar unit (logL/$L_{\odot}$); effective Temperature in Kelvin (T$_{eff}$) and Period in days (P). Only some examples are provided here. The full table is available as online material.} 
\end{table}

\begin{table}[ht]
\centering
\scriptsize
\caption{The predicted instability strip edges.} 
\label{tab:strip-edges}
\begin{tabular}{cccccc}
\hline
$M/{M_{\odot}}$ & $\log{L/{L_{\odot}}}$ &  P & ${T_{eff}}^{FBE}$ & ${T_{eff}}^{FRE}$ & \\
  & [dex] &  [days] &   K & K \\
  \hline
  &&Z= 0.0001 Y= 0.245&&\\
  \hline
  0.53 &2.453494 &3.4978121  &6400 & -\\
  0.54 &2.317279 &2.3563282  &6550 & -\\
  0.55 &2.232166 &1.9703730  &6550 & -\\
  0.56 &2.158326 &1.5512762  &6700 & -\\
  0.57 &2.095587 &1.2769337  &6750 & -\\
  0.58 &2.051199 &1.1610362  &6750 & -\\
  0.60 &1.989130 &0.9585853  &6850 & -\\
  0.62 &1.933212 &0.8422540  &6850 & -\\
  0.65 &1.866333 &0.6781071  &6950 & -\\
  0.53 &2.500738 &7.183019 & - &5350  \\
  0.54 &2.452058 &6.759199 & - &5150 \\
  0.55 &2.460422 &6.890940  &- &5050 \\
  0.56 &2.417475 &6.153294  &- &5050 \\
  0.57 &2.385255 &5.630350  &- &5050 \\
  0.58 &2.272147 &4.179636  &- &5150 \\
  0.60 &2.048857 &2.126750  &- &5550 \\
  0.62 &1.996771 &1.773650  &- &5650 \\
  0.65 &1.933041 &1.491025  &- &5650 \\
 \hline
&&Z= 0.001 Y= 0.245&&\\
 \hline
 0.51 & 2.555592 & 4.6647484 & 6317 & - \\
 0.53 & 2.206797 & 1.8559412 & 6619 &  - \\
0.56 & 1.988376 & 1.0457112 &  6808 & - \\
 0.60 & 1.819548 & 0.6666025 & 6942 & - \\
0.53 &2.411596  &6.960854 &- &4969 \\
0.55 &2.352743 &   5.907548 &- &4918\\
0.56 &2.195487 & 3.815380&  -&5108\\
0.60 &1.919358 & 1.643287 & - &5542\\
\hline
&&Z=0.01 Y=0.26&&\\
\hline
 0.495 &2.442646 &4.1462011  &6200 &-\\
 0.500 &2.272136 &2.5503497  &6389 &-\\
0.505 &2.170096 &1.9359070  &6500 &-\\
 0.510 &2.105409 &1.6513918 & 6570& -\\
 0.515 &2.018222 &1.3085711  &6650& -\\
 0.520 &1.955233 &1.1247217  &6700& -\\
 0.530 &1.844189 &0.8464146  &6800& -\\
 0.540 &1.766732 &0.6960070  &6874&-\\
 0.550 &1.652050 &0.5407182  &6907& -\\
 0.500 &2.385735 &8.335565 &-& 4850\\
 0.505 &2.375266 &8.998122 &-& 4687\\
 0.510 &2.341523 &8.333462 &-& 4650\\
 0.515 &2.312865 &7.709105 &-& 4626\\
 0.530 &1.995678 &2.863785 &-& 5100\\
 0.540 &1.926239 &2.319988 &-& 5174\\
 0.550 &1.828391 &1.668896 &-& 5382\\
\hline\\
\end{tabular}
\tablefoot{For every chemical composition, from left to right, columns list the mass, the luminosity level, the oscillation period, and the effective temperature of blue and red edge of the instability strip.}
\end{table}

\section{The new pulsation models of BL Her and W Vir} \label{sect_models}

In order to model BL Her and W Vir pulsating stars, we took into account the evolutionary predictions by \citet{bono20} for low-mass core helium burning models. As extensively discussed by several authors \citep[see e.g.][and references therein]{cs11}, along the ZAHB the helium core mass is constant, being mainly fixed by the chemical composition of the progenitor and with a negligible dependence on age for ages above a few Gyr. On the other hand, the total mass of the models decreases when the effective temperature increases from the red Horizontal Branch (HB) to extremely blue HB (EHB), in accordance with the mass lost along RGB \citep[see e.g.][]{Origlia14}. During the post-ZAHB evolution, stars with masses lower than RRLs, for a given chemical composition, enter the instability strip at higher luminosity levels, becoming BL Her or W Vir. 
In Figure ~\ref{fig:strip} (in the Appendix) we show a subset of the stellar evolutionary tracks presented by \citet{bono20} (see their Fig. 5), with a wide range of stellar masses ($M/M_{\odot}$ $\approx$ 0.50-0.90) and three different initial metal abundances, namely Z = 0.01 (top panel), Z=0.001 (middle panel), Z=0.0001 (bottom panel). The physical and numerical assumptions adopted in the computation of non-linear convective pulsation models have already been detailed in previous papers \citep[see e.g.][and references therein]{dicriscienzo07,marconi2015}. Here we only summarize the main properties of these models. They treat the nonlinear stellar pulsation including a nonlocal time-dependent treatment of convection that  adopts a free mixing length equivalent parameter to close the system of nonlinear equations. The opacity tables are the same as in \citet{DeSomma24}. In particular, the radiative Rosseland opacity is taken from the latest release of OPAL calculations \citet{ir96} for temperatures higher than $\log(T) = 4.0$ and compilations by \citet[][]{Ferguson_Alexander2005}, which include, for lower temperatures, contributions from molecules and grains. This update was performed in the context of the SPECTRUM project \citep[see][for details]{DeSomma24}. 
The nonlinear hydrodynamical equations have been integrated till a stable limit cycle is reached.
We notice that, considering the relatively high luminosity levels, only pulsation in the fundamental mode (F) is investigated for these models. Indeed, as expected on the basis of previous results \citep[see e.g][and references therein]{dicriscienzo07,marconi2015}, the region where the first overtone mode is efficient is generally limited, for a given mass, to the luminosity levels below the intersection between the first overtone and the fundamental red edge. Increasing the luminosity and decreasing the stellar mass makes this occurrence less and less probable.
Table ~\ref{table:allModels} reports the stellar parameters of computed models that are found to pulsate. The first two columns list the metal and helium abundances, the third and fourth columns the stellar mass and luminosity, and the following columns report the effective temperature and the oscillation period.


\section{Results of nonlinear model computation} \label{sect_results}

The computed pulsation models can be divided in two groups corresponding to the BL Her and W Vir classes. Following the prescriptions in the literature \citep[see][and refrences therein]{bono20}, we labelled as BL Her the models with periods shorter than about 4 days, corresponding to stellar masses from about 0.58 to about 0.65 $M_{\odot}$ and as W Vir models with periods longer than about 4 days that are less massive and brighter than the BL Her ones.

\subsection{The predicted instability strip}

The effective temperatures for the fundamental blue and red edges are reported in Table~\ref{tab:strip-edges}. The corresponding instability strips for both classes are overplotted to the three left panels of Figure ~\ref{fig:strip}, as computed for the corresponding metal abundances. In the right panels , the extrapolated RRL instability strip, as predicted by \citet{marconi2015} and shown in \citet{bono20} (green lines), is over-imposed for comparison, for each selected metallicity, together with the recently derived BL Her instability strip by \citet{Deka2024} (blue lines) and the previously computed boundaries by \citet{dicriscienzo07} (orange lines).

We notice that the new predicted instability strip is slightly wider than the extrapolated RRL one \citep{bono20} but slightly narrower than the one by \citet{Deka2024}, at least for the two lowest metal abundances. At $Z=0.01$ the predicted instability strip is moderately redder than the one by \citet{Deka2024}.
Moreover, the obtained instability strip boundaries are consistent with the results obtained from BL Her nonlinear convective pulsation models by \citet{dicriscienzo07} for the same metal abundances but using a grid of input parameters instead of following the evolutionary tracks as done in this paper. This result is expected considering that the adopted luminosity levels in \citet{dicriscienzo07} were consistent with the ones taken along the evolutionary tracks in this paper, even if here we explore higher metal abundances, smaller masses and include a space of parameters that is typical of W Vir stars.

\subsection{The predicted bolometric and multi-filter light curves}

One of the important advantages of nonlinear convective pulsation hydro-codes is the possibility to predict the light curve amplitude and morphology.
Atlas of bolometric light curves for the three selected metal abundances and the assumed stellar masses and luminosity levels were obtained and are provided electronically. An example is provided in Fig.~\ref{fig:bolcurve} in the Appendix.

We notice that the morphology of the predicted light curves significantly varies across the instability strip, with the pulsation amplitude generally decreasing with the effective temperature. However, in the case of the highest adopted metallicity, the minimum is not reached at the red boundary, but about 200 K hotter. This occurrence is similar to what is found in the case of metal-rich RRL models \citep[see e.g. fig. 4 in][]{Bono97}. Moreover, a similar behaviour was noted in previous BL Her models by \citet{dicriscienzo07}.
The physical cause of this hotter minimum has to be ascribed to the complex balance between convection and the pulsation driving mechanism in metal-rich models where iron photo-ionization produces a small but not negligible contribution to pulsation and remarkable reduction of radiative damping \citep[see][for a detailed discussion]{bonoincerpi}.
Figure~\ref{fig:velcurve}, in the Appendix, shows the radial velocity curves for the same models.
The bolometric light curves are then converted into various photometric systems, including Johnson Cousins $UBVRIJHK$, Gaia $\rm G$, $\rm G_{BP}$ and $\rm G_{RP}$, Rubin-LSST $ugrizy$ and VISTA $JY\rm K_s$ bands.
In Fig.~\ref{fig:gaiacurve} and ~\ref{fig:lsstcurve}, in the Appendix, we show the same curves already displayed in Fig.~\ref{fig:bolcurve} but in the Gaia and Rubin-LSST bands, respectively.

These plots confirm the trends for the pulsation amplitude and morphology as a function of the effective temperature, already noted for the bolometric curves.
We remark that we are providing the first theoretical light curves of T2Cs in the Gaia  filters.

\subsection{The multi-filter period-amplitude diagrams}

From the transformed light curves, mean magnitudes, colours and pulsation amplitudes can be derived for any selected photometric system.
In Fig.~\ref{fig:gaiabailey}, in the Appendix, we show the predicted period-amplitude diagram in the three Gaia bands for all the computed models and the labelled metal abundances. The same diagram but for the Rubin-LSST bands is shown in Fig.~\ref {fig:lsstbailey}. In both plots, models are colour-coded by the assumed stellar mass.
The well-known property of the pulsation amplitude decreasing as the wavelength increases is particularly evident in Fig.~\ref {fig:lsstbailey}, in the Appendix, where the maximum amplitude decreases from around 2 mag in the $u$ band to around 1 mag in the $z$ and $y$ filters, independently of the adopted metal content.

\section{The new theoretical relations} \label{sect_modeled_relations}

\subsection{The Period-Luminosity-Mass-Temperature-Metallicity relations}

Similarly to what was performed for classical Cepheids and RRL in our previous papers \citep{desomma2022ApJS..262...25D,marconi2015}, the derived pulsation periods (reported in Table~\ref{table:allModels}) were used in combination with the model mass, luminosity, effective temperatures and metallicities to derive the following linear Period-Luminosity-Mass-Temperature-Metallicity ($PLMTZ$) relation, also known as van Albada-Baker relation \citep{van1971masses}:
\begin{multline}
     \rm \log(P) =  (10.85\pm 0.11)+(0.924\pm0.009)\cdot \log(L/L_\odot) +\\
     -(0.677\pm 0.081)\cdot \log(M/M_\odot) - (3.338 \pm 0.029)\cdot \log(T_{eff}) +\\
     (0.015 \pm 0.002)\cdot \log(Z)
\end{multline}
characterized by an $rms=0.014$ dex and a determination coefficient $\rm R^2=0.99$\footnote{The coefficient of determination, $R^2$, quantifies the proportion of variance in the observed data that is explained by the model. An $R^2$ value close to 1 indicates a good fit, while values near 0 imply a poor explanatory power.}
The coefficient of the metallicity term $[Fe/H]$ is found to be generally below $0.1$ mag/dex with a sign in agreement with what has already been found for RRLs \citep{marconi2015,marconi2018impact,marconi2022}. This metallicity dependence is small but has to be taken into account to apply these relations to derive accurate pulsation periods.

\subsection{The Period-Luminosity relations}

\begin{figure*}[ht]
    \includegraphics[trim=0cm 2.7cm 0cm 0cm, clip, width=1.0\linewidth]{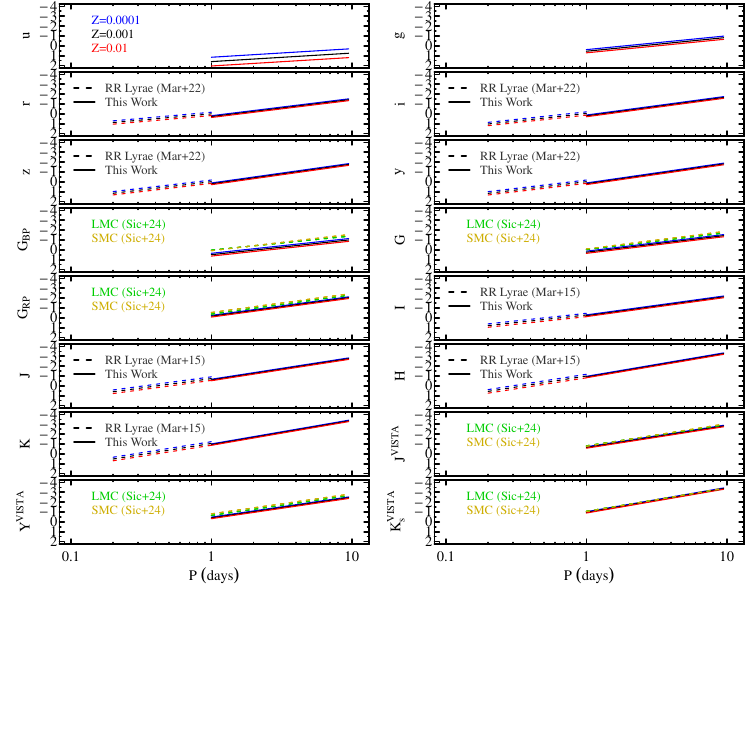}
    \caption{PL relations obtained for several bands. In order, from top to bottom and left to right, the Gaia $G_{BP}, G, G_{RP}$, the Johnson-Cousinhs $I, J, H, K$, the VISTA $J, Y, K_s$ and the Rubin-LSST $g,r, i,z,y$ bands. Different colours reflect the three different metallicities. The dashed lines correspond to the relations from RRL models \citep{marconi2015, marconi2022}, while the relations from \citet{sicignano24} for the Gaia and VISTA bands are plotted in green and yellow.}
    \label{fig:PL_relations}
\end{figure*}

The obtained PL relations in the Gaia $G_{BP}, G, G_{RP}$, Johnson-Cousinhs $I, J, H, K$, the VISTA $J, Y, K_s$ bands and LSST $g,r, i,z,y$ bands are shown in Fig.~\ref{fig:PL_relations} for the three assumed metal-abundances. Coefficients results for all the considered bands are listed in Table ~\ref{tab:PW_relations_withRrl_no_u_band_with_PL_VISTA_edgeErr_coeffs}.

\begin{table*}[ht]
\centering
\caption{The coefficients of the $PLZ$ and $PWZ$ relations.} 
\footnotesize\setlength{\tabcolsep}{3.0pt}
\begin{tabular}{lllllccccc}
\hline
$\alpha$ & $\beta$ & $\gamma$ & rms & $R^2$ & bands &  $\rm \mu_0^{LMC}$ & $\rm N_{dat}^{LMC}$ & $\rm \mu_0^{SMC}$ & $\rm N_{dat}^{SMC}$ \\ 
\hline
\hline
    0.672 $\pm$ 0.019 & -1.524 $\pm$ 0.029 & 0.1409 $\pm$ 0.01 & 0.281 & 0.76 & $\rm G_{BP}$ & - & - & - & - \\ 
  0.385 $\pm$ 0.015 & -1.716 $\pm$ 0.023 & 0.1106 $\pm$ 0.0083 & 0.233 & 0.85 & G  & - & - & - & - \\ 
  -0.056 $\pm$ 0.011 & -1.92 $\pm$ 0.019 & 0.0859 $\pm$ 0.0064 & 0.183 & 0.92 & $\rm G_{RP}$ & - & - & - & - \\ 
  -0.101 $\pm$ 0.011 & -1.955 $\pm$ 0.018 & 0.0865 $\pm$ 0.0059 & 0.174 & 0.93 & I & - & - & - & - \\ 
  -0.4976 $\pm$ 0.0071 & -2.207 $\pm$ 0.012 & 0.0755 $\pm$ 0.0043 & 0.119 & 0.97 & J & - & - & - & - \\ 
  -0.8003 $\pm$ 0.0041 & -2.4264 $\pm$ 0.0079 & 0.0711 $\pm$ 0.0027 & 0.073 & 0.99 & H & - & - & - & - \\ 
  -0.8435 $\pm$ 0.0038 & -2.4546 $\pm$ 0.0074 & 0.0707 $\pm$ 0.0024 & 0.069 & 0.99 & K & - & - & - & - \\ 
  -0.5506 $\pm$ 0.0068 & -2.22 $\pm$ 0.011 & 0.0741 $\pm$ 0.0043 & 0.116 & 0.97 & $\rm J^{vista}$ & - & - & - & - \\ 
  -0.3016 $\pm$ 0.0094 & -2.091 $\pm$ 0.014 & 0.0778 $\pm$ 0.0053 & 0.146 & 0.96 & $\rm Y^{vista}$ & - & - & - & - \\ 
  -0.8738 $\pm$ 0.0041 & -2.4612 $\pm$ 0.0073 & 0.0669 $\pm$ 0.0026 & 0.068 & 0.99 & $\rm K_s^{vista}$ & - & - & - & - \\ 
   0.442$\pm$ 0.015 & -0.864$\pm$ 0.050 & 2.170$\pm$ 0.026 & 0.439& 0.51 & $\rm u_{lsst}$  & - & - & - & - \\ 
  0.755 $\pm$ 0.021 & -1.397 $\pm$ 0.031 & 0.162 $\pm$ 0.012 & 0.316 & 0.68 & $\rm g_{lsst}$ & - & - & - & - \\ 
  0.395 $\pm$ 0.014 & -1.744 $\pm$ 0.022 & 0.0902 $\pm$ 0.0081 & 0.226 & 0.86 & $\rm r_{lsst}$ & - & - & - & - \\ 
  0.317 $\pm$ 0.012 & -1.906 $\pm$ 0.019 & 0.0859 $\pm$ 0.0066 & 0.186 & 0.92 & $\rm i_{lsst}$ & - & - & - & - \\ 
  0.312 $\pm$ 0.01 & -2.003 $\pm$ 0.016 & 0.0866 $\pm$ 0.0055 & 0.165 & 0.94 & $\rm z_{lsst}$ & - & - & - & - \\ 
  0.3015 $\pm$ 0.0094 & -2.051 $\pm$ 0.016 & 0.0835 $\pm$ 0.005 & 0.156 & 0.95 & $\rm y_{lsst}$ & - & - & - & - \\ 
-0.6135 $\pm$ 0.0041 & -2.7135 $\pm$ 0.0086 & -0.1119 $\pm$ 0.003 & 0.073 & 0.99 & $\rm rgr_{lsst}$ & - & - & - & - \\ 
  -0.2459 $\pm$ 0.0027 & -2.5608 $\pm$ 0.005 & -0.0127 $\pm$ 0.0019 & 0.046 & 1.00 & $\rm igi_{lsst}$ & - & - & - & - \\ 
  0.2958 $\pm$ 0.0057 & -2.3149 $\pm$ 0.01 & 0.089 $\pm$ 0.0035 & 0.099 & 0.98 & $\rm ziz_{lsst}$ & - & - & - & - \\ 
  0.0476 $\pm$ 0.0043 & -2.4168 $\pm$ 0.0068 & 0.0393 $\pm$ 0.0026 & 0.072 & 0.99 & $\rm ygy_{lsst}$ & - & - & - & - \\ 
  -1.1254 $\pm$ 0.0028 & -2.5255 $\pm$ 0.0051 & 0.0402 $\pm$ 0.0018 & 0.047 & 1.00 & IVI & 18.44(0.11) & 149 & 18.80(0.31) & 26 \\ 
  -0.9046 $\pm$ 0.0039 & -2.4457 $\pm$ 0.0066 & 0.0598 $\pm$ 0.0023 & 0.067 & 0.99 & JVJ & - & - & - & - \\ 
  -0.8329 $\pm$ 0.0044 & -2.4199 $\pm$ 0.0078 & 0.0662 $\pm$ 0.0026 & 0.074 & 0.99 & JIJ & - & - & - & - \\ 
  -1.0949 $\pm$ 0.0027 & -2.6084 $\pm$ 0.0056 & 0.0613 $\pm$ 0.0018 & 0.044 & 1.00 & HVH & - & - & - & - \\ 
  -1.0901 $\pm$ 0.0029 & -2.6216 $\pm$ 0.0055 & 0.0647 $\pm$ 0.0019 & 0.044 & 1.00 & HIH & - & - & - & - \\ 
  -1.3373 $\pm$ 0.0041 & -2.8156 $\pm$ 0.0071 & 0.0632 $\pm$ 0.0022 & 0.052 & 1.00 & HJH & - & - & - & - \\ 
  -1.0243 $\pm$ 0.0029 & -2.5665 $\pm$ 0.0059 & 0.0649 $\pm$ 0.0018 & 0.049 & 1.00 & KVK & - & - & - & - \\ 
  -1.0153 $\pm$ 0.003 & -2.5701 $\pm$ 0.006 & 0.0671 $\pm$ 0.0019 & 0.050 & 1.00 & KIK & - & - & - & - \\ 
  -1.084 $\pm$ 0.0027 & -2.6267 $\pm$ 0.0062 & 0.0674 $\pm$ 0.0017 & 0.044 & 1.00 & KJK & - & - & - & - \\ 
  -0.9207 $\pm$ 0.0036 & -2.5049 $\pm$ 0.0065 & 0.0702 $\pm$ 0.0023 & 0.061 & 0.99 & KHK & - & - & - & - \\ 
  -0.9669 $\pm$ 0.0034 & -2.4682 $\pm$ 0.0061 & 0.0543 $\pm$ 0.0022 & 0.060 & 0.99 & JVI & - & - & - & - \\ 
  -1.1004 $\pm$ 0.0026 & -2.5935 $\pm$ 0.0051 & 0.0575 $\pm$ 0.0018 & 0.043 & 1.00 & HVI & - & - & - & - \\ 
  -1.036 $\pm$ 0.0028 & -2.5617 $\pm$ 0.0057 & 0.062 $\pm$ 0.002 & 0.049 & 1.00 & KVI & - & - & - & - \\ 
  -0.9917 $\pm$ 0.0032 & -2.4655 $\pm$ 0.0055 & 0.0065 $\pm$ 0.0022 & 0.059 & 0.99 & $\rm GG_{BP}G_{RP}$ & 18.48(0.13) & 149 & 18.89(0.26) & 26 \\ 
  -1.1208 $\pm$ 0.0029 & -2.6456 $\pm$ 0.0062 & 0.0614 $\pm$ 0.0016 & 0.044 & 1.00 & $\rm (K_sJK_s)^{vista}$ & 18.45(0.13) & 149 & 18.82(0.32) & 26 \\ 
  -1.1425 $\pm$ 0.0027 & -2.6349 $\pm$ 0.0059 & 0.0618 $\pm$ 0.0018 & 0.044 & 1.00 & $\rm (K_sYK_s)^{vista}$& 18.47(0.12) & 149 & 18.87(0.34) & 26 \\ 
   \hline
\end{tabular}
\tablefoot{The coefficients of the predicted relations $mag = \alpha + \beta \log P + \gamma [\mathrm{Fe/H}]$, fitted for different band combinations and 886 models, are listed in columns 1–3. Column 4 reports the root mean square ($rms$) of the residuals around the fit, and column 5 contains the coefficient of determination $R^2$. Column 6 indicates the band or the band combination. The last four columns present the median (with robust standard deviation) of the distance distributions obtained by applying the fitted relation to a sample of T2Cs belonging to the LMC and SMC.}
\label{tab:PW_relations_withRrl_no_u_band_with_PL_VISTA_edgeErr_coeffs}
\end{table*}

In Fig. \ref{fig:PL_relations}, the relations obtained for RRL models \citep{marconi2015,marconi2022} and the empirical ones obtained from VISTA data by \citet{sicignano24}, for the corresponding metallicities and filters, are overlapped. It seems quite evident that: i) the predicted metallicity effect on the zero point of the NIR PL relations is generally smaller for T2Cs than for RRLs, but the slopes are highly consistent; ii) we find an excellent agreement with results by \citet{sicignano24}.

We note that an often overlooked source of uncertainty arises when T2Cs stars are inadvertently classified as RRL (or vice versa) and the corresponding PL or PLW relations are applied. Based on our results (see Figs. 8 and 9), this effect is negligible at higher metallicities, where the PL relations of the two classes nearly overlap. However, at lower metallicities, the PL zero-points for T2Cs are systematically fainter, implying that using RRL PL relations would lead to brighter absolute magnitudes and consequently to underestimated distances by some tenths of magnitudes.)

\subsection{The Period-Wesenheit relations}

\begin{figure*}
  \centering
\includegraphics[trim=0cm 0.5cm 0cm 0cm, clip, width=\textwidth]{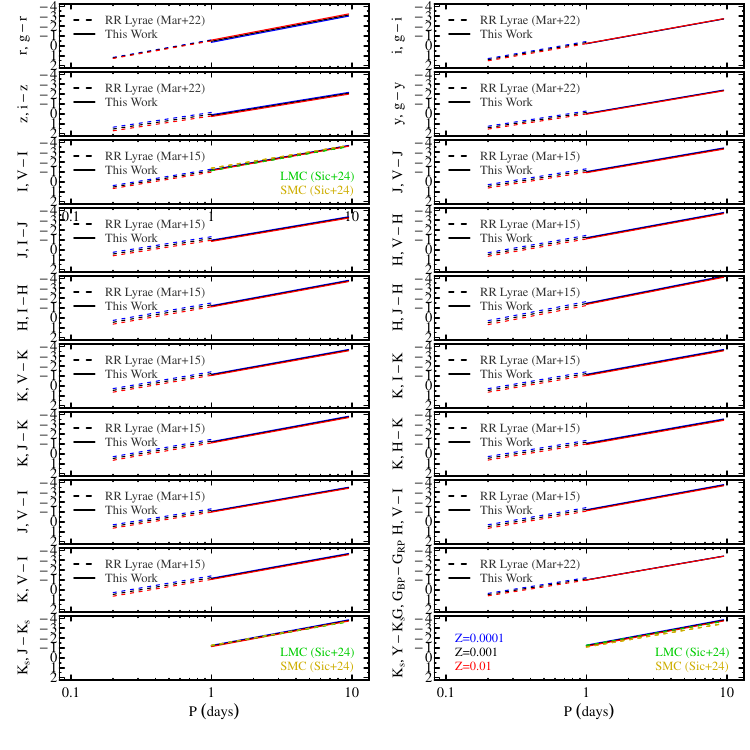}
   \caption{PW relation calculated in this work. Colours and line-style are the same as Fig.~\ref{fig:PL_relations}}\label{fig:PW_relations}
\end{figure*}

In Fig.~\ref{fig:PW_relations}, we show the obtained PW relations\footnote{Wesenheit magnitudes, introduced by \citet{Madore1982}, provide
reddening-free magnitudes once we assume to know the extinction law. They are defined as $W_{X_1,X_2-X_3} = X_1 -\xi^{2;3}_1\cdot(X_2 - X_3)$, where $X_i$ indicates the generic band and the coefficient $\xi^{2,3}_1$ coincides with the total-to-selective absorption.} for the selected metal abundances and the labelled filter combinations. Again, theoretical RRL relations from \citet{marconi2015} are overimposed where available, and a comparison with the empirical results by \citet{sicignano24} is shown for the VISTA band combinations.
Similarly to what was found for RRL \citep[see e.g.][for detail]{marconi2015,marconi2022}, we identify optical band combinations that minimise the metallicity effect.
In particular, the metallicity dependence seems to be completely negligible in the case of the PW relation involving the Rubin-LSST $g$ and $i$ bands and very small (but still significant given the tiny errors on coefficients), in all the other cases, making these relations very solid tools to derive individual T2C distances irrespective of uncertainties in their metal abundance.
These results are in agreement with the empirical evidence provided by \citet{GJ17} who found no difference in the PL relations of T2Cs in the LMC and SMC.

\section{Application to observed T2Cs} \label{sect_observed_relations}

As an application of the previously derived relationships, we calculated the distances of various samples of T2Cs collected from the literature. Specifically, in the following sections, we will focus on datasets belonging to the Large Magellanic Cloud (LMC), the Small Magellanic Cloud (SMC), and the Milky Way (MW).

\subsection{LMC and SMC distances}
We used two samples of 149 LMC and 26 SMC T2Cs, respectively, with photometric data available from Gaia Data Release 3 
\citep[DR3][]{GaiaVallenari2023,Ripepi2023}. The adopted mean values of [Fe/H] are $-0.409\pm0.076$ dex and $-0.78\pm0.08$ dex, for the LMC and the SMC, respectively, are taken from the literature \citep[see e.g.][]{Romaniello2008A&A...488..731R, Romaniello2022A&A...658A..29R}.

The Wesenheit magnitude in the Gaia bands, defined as $\rm W_{Gaia}=G - 1.9\cdot (G_{BP}-G_{RP})$ \citep{Ripepi2019A&A...625A..14R}, was taken into account. The observed Wesenheit magnitudes were obtained by applying the last equation to the mean magnitudes in the $\rm G$, $\rm G_{BP}$ and $\rm G_{RP}$, retrieved from the public Gaia archive.

The absolute Wesenheit magnitudes in the Gaia photometric bands were obtained by applying the corresponding theoretical $PWZ$ relation (reported in Table~\ref{tab:PW_relations_withRrl_no_u_band_with_PL_VISTA_edgeErr_coeffs}) to all the selected sources.
From the difference between the absolute and observed Wesenheit magnitudes, we infer the individual distance moduli ($\mu_0$).
The associated errors are evaluated by taking into account both the observational uncertainties and the $rms$ of the adopted relations. 

The best distance moduli for the LMC and SMC (columns 8 and 10 of Table~\ref{tab:PW_relations_withRrl_no_u_band_with_PL_VISTA_edgeErr_coeffs} were determined by computing the median of the distributions of the individual $\mu_0$ values derived for the T2Cs described above. The uncertainties associated with these best estimates were calculated using the robust standard deviation, defined as 1.4826$\cdot\mathrm{MAD}$, of the corresponding distributions.

\begin{figure*}[ht]
    \includegraphics[trim=0cm 0cm 0cm 0.0cm, clip, width=1.0\linewidth]{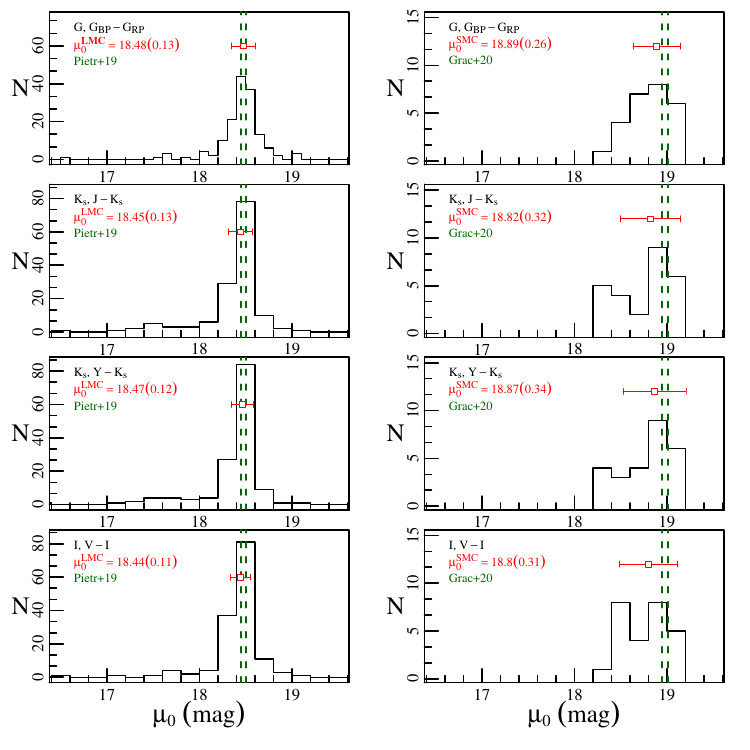}
    \caption{Distributions of the $\mu_0$ values obtained by applying our $PWZ$ theoretical relations to a sample of LMC T2Cs (left panels) and SMC T2Cs (right panels). The results for different bands are plotted in different panels. From top to bottom: the Gaia $\rm G,BP-RP$, the VISTA $\rm K_s,J-K_s$, the VISTA $\rm K_s,Y-K_s$, and the Johnson $\rm I,V-I$ $PWZ$ relations. The vertical green lines represent the LMC and SMC distance ranges around the currently adopted geometric best values from \citet{Pietrzynski_2019Natur.567..200P} and \citet{Graczyk_2020ApJ...904...13G}, respectively, while the red square and the error bar indicate the median value of our distribution and its robust standard deviation (i.e. 1.4826$\cdot\mathrm{MAD}$), respectively.}
    \label{fig:mu0_mc}
\end{figure*}

In Fig.~\ref{fig:mu0_mc}, we show the distribution of $\mu_0$ values obtained by applying our theoretical $PWZ$ relations to the selected LMC (left column) and SMC (right column) T2Cs. The results for different bands are plotted in different panels. We show that our results are slightly shorter than the value by \citet{Pietrzynski_2019Natur.567..200P} and \citet{Graczyk_2020ApJ...904...13G} but still consistent within the errors. The worse quality of the inferred distribution in the case of the SMC is likely due to the small number of considered Cepheids.

\subsection{Galactic T2Cs}

We also selected a sample of 100 Galactic T2Cs from the Gaia DR3 as detailed in \citet{sicignano24} and repeated the analysis described above, adopting the $PWZ$ relation in the Gaia photometric bands. Since metallicity measurements for these stars are unavailable, we tested our $PWZ$ relations by assuming three fixed metallicity values, namely $Z = 0.01$, $Z = 0.001$, and $Z = 0.0001$, which correspond to those used in the computation of the pulsation models.

\begin{figure*}[ht]
    \includegraphics[trim=0cm 0cm 0cm 0cm, clip, width=1.0\linewidth]{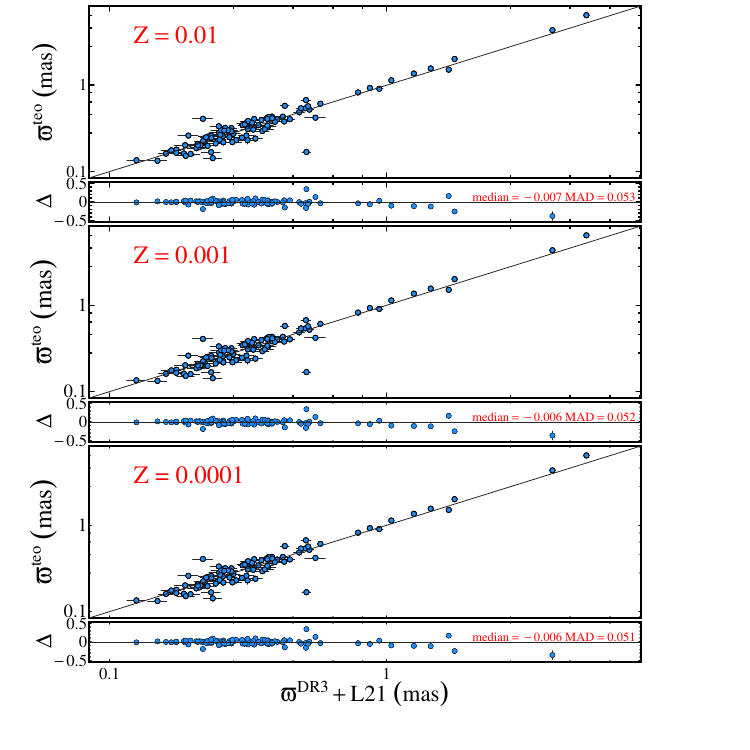}
    \caption{Theoretical parallaxes of Galactic Type II Cepheids in the Gaia database, as obtained from inversion of the theoretical $PWZ$ relation in the Gaia bands, versus Gaia astrometric values, as corrected for the \citet{lindegren_2021A&A...649A...4L} offset. In each panel, we show the results obtained assuming the labelled metal abundance in the $PWZ$ relation. The bottom plot of each panel shows the corresponding residuals computed as $\rm \varpi^{DR3}+L21 - \varpi^{teo}$. \label{fig:parallax_comparison} }
\end{figure*}

The individual $\mu_0$, derived by applying the $PWZ$ relation, was converted into parallaxes to enable a direct comparison with Gaia DR3 results, including the zero-point correction provided by \citet{lindegren_2021A&A...649A...4L} (L21). The resulting theoretical parallaxes ($\varpi^{\mathrm{teo}}$), along with their associated uncertainties, are reported in Table~\ref{tab:mw-parallaxes}, together with the Gaia DR3 parallaxes and L21 correction values. The comparison with Gaia astrometric results is shown in Fig.~\ref{fig:parallax_comparison}, for the three assumed metallicities.
In each panel, the predicted theoretical parallaxes are plotted against the corrected Gaia DR3 values with a 1:1 reference line shown to make the comparison easier. The residuals, computed as $\varpi^{\mathrm{DR3}} + \mathrm{L21} - \varpi^{\mathrm{teo}}$, are displayed in the three smaller panels, centred around the zero line. A statistical analysis of these residuals indicates that Gaia DR3 parallaxes, after applying the L21 correction, are systematically larger than those inferred from the pulsation models. The median differences range from a marginally significant value of $\rm 9 \pm 5\ \mu as$ for the most metal-poor models, up to $\rm 14.0 \pm 5.4\ \mu as$ for the most metal-rich ones. The uncertainties on the median values are estimated as $\mathrm{MAD} / \sqrt{100}$. Our results are consistent with recent findings in the literature, which suggest that the application of the correction proposed by \citet{lindegren_2021A&A...649A...4L} tends to over-correct Gaia DR3 parallaxes, resulting in values that are systematically larger than expected. This trend has been noted in several independent studies \citep[see, e.g.,][for a recent review]{molinaro_2023MNRAS.520.4154M}.

\begin{table}[ht]
\centering
\caption{Adopted Galactic T2Cs.} 
\label{tab:mw-parallaxes}
\footnotesize\setlength{\tabcolsep}{3.0pt}
\begin{tabular}{cccccccc}
  \hline
ID & class & P & $\rm L21_{corr}$ & $\rm \varpi_{DR3}$ & $\varpi^{teo}_{Z0.0001}$ & $\varpi^{teo}_{Z0.001}$ & $\varpi^{teo}_{Z0.01}$ \\ 
  \hline
DR2\_2026857124982231296 & BLHER & 1.0177 & -0.041 & 0.292 $\pm$ 0.023 & 0.405 $\pm$ 0.012 & 0.406 $\pm$ 0.012 & 0.408 $\pm$ 0.012 \\ 
  DR2\_1816085861226864768 & BLHER & 1.0918 & -0.021 & 0.365 $\pm$ 0.015 & 0.43 $\pm$ 0.012 & 0.432 $\pm$ 0.012 & 0.433 $\pm$ 0.012 \\ 
  ATLAS\_J085.3704+33.6699 & BLHER & 1.1037 & -0.042 & 0.215 $\pm$ 0.02 & 0.2172 $\pm$ 0.0059 & 0.2179 $\pm$ 0.0059 & 0.2185 $\pm$ 0.0059 \\ 
  ASAS\_J140159-4256.3 & BLHER & 1.1525 & -0.028 & 0.358 $\pm$ 0.017 & 0.382 $\pm$ 0.012 & 0.383 $\pm$ 0.012 & 0.385 $\pm$ 0.012 \\ 
  ZTFJ181910.47+033444.7 & BLHER & 1.1630 & -0.039 & 0.21 $\pm$ 0.02 & 0.2827 $\pm$ 0.0077 & 0.2835 $\pm$ 0.0077 & 0.2844 $\pm$ 0.0077 \\ 
  DR2\_5431999418776812160 & BLHER & 1.2017 & -0.037 & 0.234 $\pm$ 0.016 & 0.2486 $\pm$ 0.0068 & 0.2494 $\pm$ 0.0068 & 0.2501 $\pm$ 0.0068 \\ 
  DR2\_5784674881551467392 & BLHER & 1.2380 & -0.016 & 0.36 $\pm$ 0.014 & 0.427 $\pm$ 0.012 & 0.429 $\pm$ 0.012 & 0.43 $\pm$ 0.012 \\ 
  ...................\\
  ROTSE1 J180955.16+235746.7 & WVIR & 12.5634 & -0.013 & 0.1607 $\pm$ 0.0093 & 0.1645 $\pm$ 0.0046 & 0.165 $\pm$ 0.0046 & 0.1655 $\pm$ 0.0046 \\ 
  DR2\_532154183216635520 & WVIR & 12.6450 & -0.016 & 0.337 $\pm$ 0.012 & 0.394 $\pm$ 0.011 & 0.395 $\pm$ 0.011 & 0.396 $\pm$ 0.011 \\ 
  COMP\_VAR\_VSX\_2019 & WVIR & 12.7085 & -0.043 & 0.471 $\pm$ 0.019 & 0.1679 $\pm$ 0.0055 & 0.1684 $\pm$ 0.0055 & 0.1689 $\pm$ 0.0055 \\ 
  WY\_Sco & WVIR & 12.9247 & -0.016 & 0.192 $\pm$ 0.017 & 0.2 $\pm$ 0.0058 & 0.2006 $\pm$ 0.0058 & 0.2012 $\pm$ 0.0058 \\ 
  DR2\_2037889315421818496 & WVIR & 12.9637 & -0.033 & 0.226 $\pm$ 0.022 & 0.297 $\pm$ 0.01 & 0.298 $\pm$ 0.01 & 0.299 $\pm$ 0.01 \\ 
   \hline
\end{tabular}
 \tablefoot{These targets used for testing the distances obtained by applying our $PWZ$ relations. The full table is available as online material.}
\end{table}

\subsection{Globular Clusters}
\begin{figure*}[ht]
    \includegraphics[trim=0cm 0cm 0cm 0cm, clip, width=1.0\linewidth]{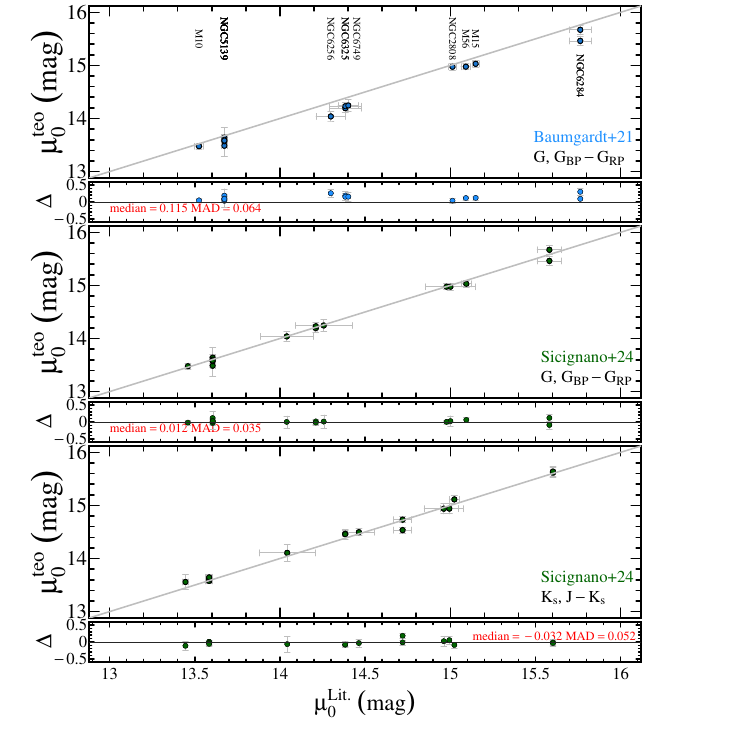}
    \caption{Distance moduli derived in this work (vertical axis) for GGC T2Cs pulsators compared to those from \citet{Baumgardt2021} in the Gaia bands (top large panel), from \citet{sicignano24} in the Gaia bands (middle large panel), and \citet{sicignano24} in the $\rm K_s, J - K_s$ bands (bottom large panel). The residuals are shown in the corresponding smaller panels. \label{fig:mu0_GC} }
\end{figure*}

As a further test, we applied our PWZ relations in the Gaia bands to a sample of T2C host GGCs. The three large panels of Fig.~\ref{fig:mu0_GC} display our derived distance moduli (vertical axis) compared with reference values from the literature: \citet{Baumgardt2021} in the top panel, and \citet{sicignano24} in the middle and bottom panels. In each case, the reference values are plotted on the horizontal axis. The corresponding residuals to the 1:1 relation are shown in the smaller panels below each main plot.
A quick inspection suggests that \citet{sicignano24} distance moduli are in better agreement with the theoretical predictions than the value published by \citet{Baumgardt2021}. This is not a surprise as \citet{sicignano24} using their derived empirical calibrations of PL and PW relations obtained distances to these T2C-host GGCs systematically smaller by $\sim$ 0.1 mag and 0.03$-$0.06 mag than \citet{Baumgardt2021} when the zero points are calibrated with the distance of the LMC or Gaia parallaxes, respectively. Moreover, it is worth noting that \citet{bhardwaj2023precise} also found that their derived RRL-based GGCs distances are systematically smaller, by about 0.015 mag, than those provided by \citet{Baumgardt2021}.

\section{Conclusions} \label{sect_conclusions}

A new theoretical scenario was presented for BL Her and W Vir pulsating stars with periods up to about 10 days for three selected metal abundances (Z= 0.01, 0.001 and 0.0001).
The main results of this investigation are reported below
\begin{itemize}

\item The obtained model predictions show that the new T2Cs theoretical instability strip is consistent with the results obtained from BL Her nonlinear convective pulsation models by \citet{dicriscienzo07}, slightly wider than the extrapolated RRL one \citep{marconi2015}, but slightly narrower and/or redder than the one by \citet{Deka2024}, depending on the assumed metal content.

\item The morphology of the predicted light curve varies across the instability strip with similarities to the case of RRL stars. Through the adoption of model atmospheres, we obtained theoretical light curves and pulsation amplitudes in a variety of photometric systems and for the first time in the Gaia and Rubin-LSST filters for T2C models.

\item The obtained mean magnitudes and colours have been adopted to build theoretical PL and PW relations in all the considered photometric bands and to investigate the effect of the assumed metal abundance. We have found that the predicted metallicity effect on the zero point of the NIR PL relations is generally
smaller for T2Cs than for RRL. Moreover, we find an 
excellent agreement with the results by \citet{sicignano24}.
In the case of the PW relations, similarly to
what was found for RRL, we identified optical band combinations that minimise the metallicity dependence. 

\end{itemize}

The application of the inferred theoretical relations to Magellanic Cloud and Galactic T2C data shows a good agreement. In particular, the LMC and SMC distance moduli $\mu_0$ obtained as the medians of the distributions of the individual values derived
for the investigated T2Cs, are found to be slightly shorter than the value by \citet{Pietrzynski_2019Natur.567..200P} and \citet{Graczyk_2020ApJ...904...13G}, but still consistent within the errors. 
As for Galactic T2Cs, 100 targets taken from the ESA Gaia DR3 were considered. The individual distance moduli, derived by applying the 
PWZ relations were converted into parallaxes to enable a direct comparison with Gaia DR3 values, including the zero-point correction provided by \citet{lindegren_2021A&A...649A...4L}.
As a result, we found a deviation of theoretical parallaxes from Gaia results, which is consistent with published independent indications that \citet{lindegren_2021A&A...649A...4L} correction tends to overestimate Gaia parallaxes \citep[see also][for details]{molinaro_2023MNRAS.520.4154M}.
We finally applied our PWZ relations in the Gaia bands to a sample of T2C host GGCs. As a result, we found that \citet{sicignano24} distance moduli are in better agreement with the theoretical predictions than the value published by \citet{Baumgardt2021}.
On the basis of the obtained relations and distances, we can conclude that T2C nonlinear convective pulsation models provide additional tools to constrain individual distances and stellar properties of the host old stellar populations. 
In the future we plan to extend the model set to a wider range of input parameters, to apply the model-fitting technique \citep[see e.g.][and references therein]{Molinaro25}, to reproduce the observed multi-filter light curves available in current and future surveys and to provide a theoretical calibration of the Tip of the RGB as an alternative population II distance indicator calibrating SNIa in the stellar route to the Hubble constant \citep[see e.g.][and references therein]{Freedman21}.

\section{Data availability}  
Tables 1 and 4 are only available in electronic form at the CDS via anonymous ftp to cdsarc.u-strasbg.fr (130.79.128.5) or via http://cdsweb.u-strasbg.fr/cgi-bin/qcat?J/A+A/.

\begin{acknowledgements} 
We thank the anonymous referee for her/his very useful comments.
We acknowledge the financial support from INAF (Large Grant MOVIE, PI: Marconi), Project PRIN MUR 2022 (code 2022ARWP9C) “Early Formation and Evolution of Bulge and HalO (EFEBHO)", PI: M. Marconi, funded by European Union – Next Generation EU, ASI-Gaia (“Missione Gaia Partecipazione italiana al DPAC – Operazioni e Attività di Analisi dati”), the INAF GO-GTO grant 2023 "C-MetaLL – Cepheid Metallicities in the Leavitt Law" (PI: V. Ripepi) and INAF-ASTROFIT fellowship (PI: G. De Somma). 
This research was also supported by the International Space Science Institute (ISSI) in Bern, through ISSI International Team project SHoT: The Stellar Path to the Ho Tension in the Gaia, TESS, LSST and JWST Era.
G.D.S. and T.S. thank INFN (Naples section) for support via QGSKY initiatives, with additional INFN support for Moonlight2 (G.D.S.).
This research also benefited from COST Action CA21136 (CosmoVerse), addressing cosmological tensions through systematics and fundamental physics, funded by COST (European Cooperation in Science and Technology).
\end{acknowledgements}

\bibliographystyle{aa}
\bibliography{aa56458-25.bib}

\begin{appendix}
\section{Figures that are not shown in the text}  

\begin{figure*}[ht]
    \hbox{
    \includegraphics[trim=0cm 0.24cm 0cm 0.0cm, clip, width=0.507\linewidth]{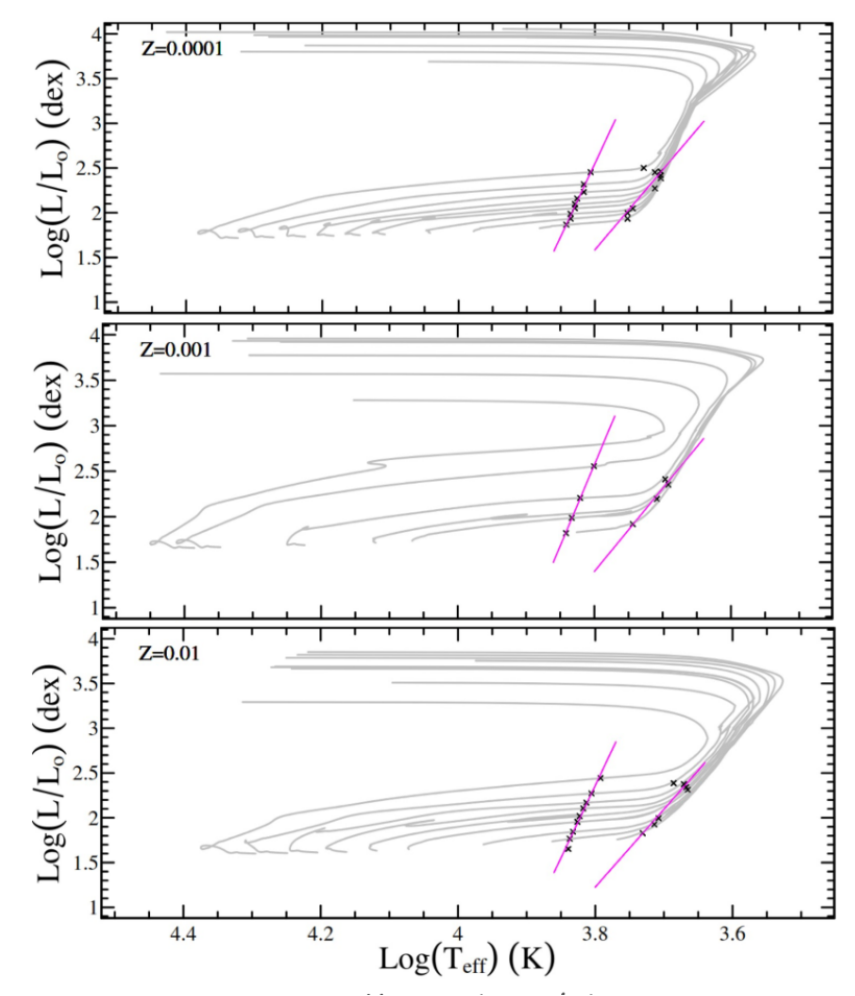}
    \includegraphics[trim=0cm 9cm 0cm 0.0cm, clip,width=0.485\linewidth]{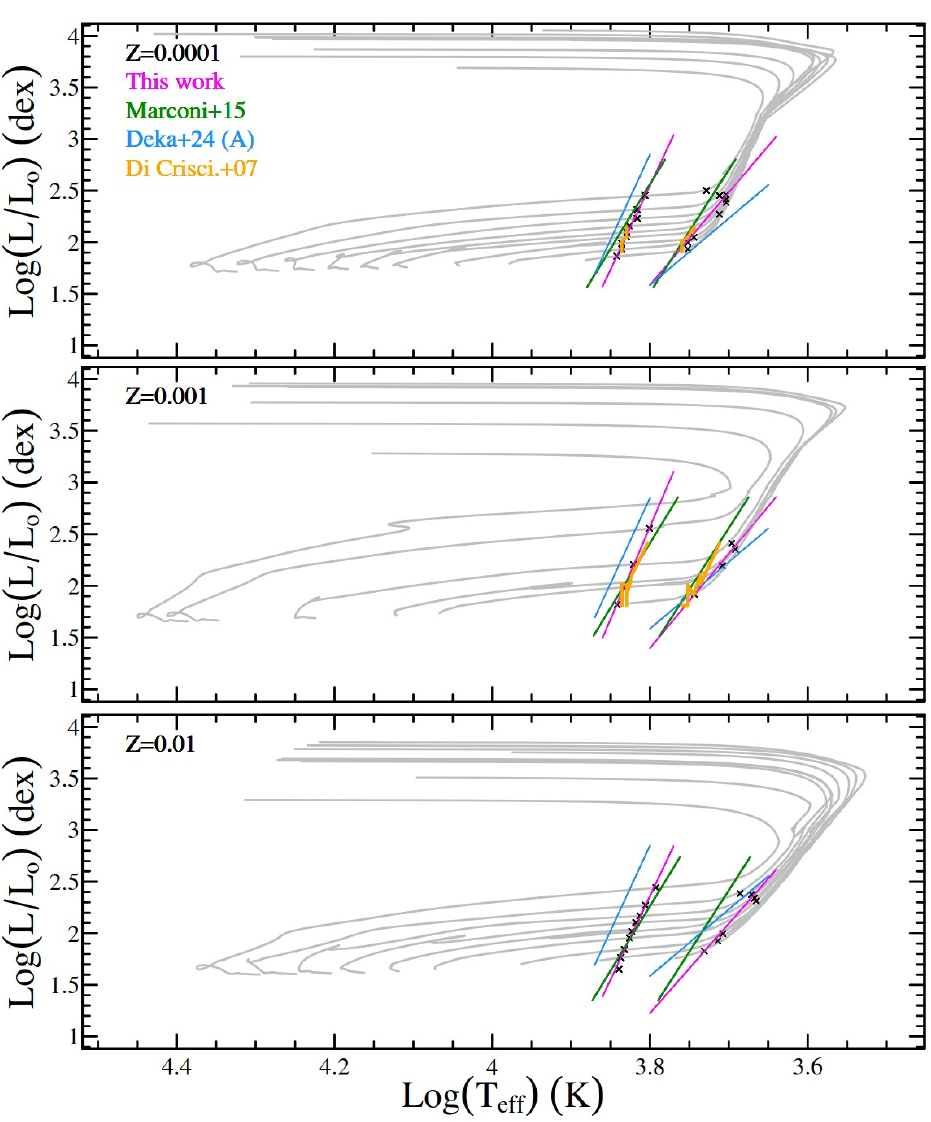}
    }
    \caption{\label{fig:strip} Left: the predicted F-mode instability strips (solid lines) for the three labelled metallicities over-imposed to the evolutionary tracks from \citet{bono20}. Right: the same but compared with the extrapolated RRL instability strip, as predicted by \citet{marconi2015} and shown in \citet{bono20} (green lines), for each selected metallicity, together with the recently derived BL Her instability strip by \citet{Deka2024} (blue lines) and the previously computed boundaries by \citet{dicriscienzo07} (orange lines).}

\end{figure*}

In Figure ~\ref{fig:strip} we show a subset of the stellar evolutionary tracks presented by \citet{bono20} (see their Fig. 5), with a wide range of stellar masses ($M/M_{\odot}$ $\approx$ 0.50-0.90) and three different initial metal abundances, namely Z = 0.01 (top panel), Z=0.001 (middle panel), Z=0.0001 (bottom panel). 

\begin{figure*}[ht]
    \includegraphics[trim=2.4cm 0.0cm 0cm 0.0cm, clip, width=1.0\linewidth]{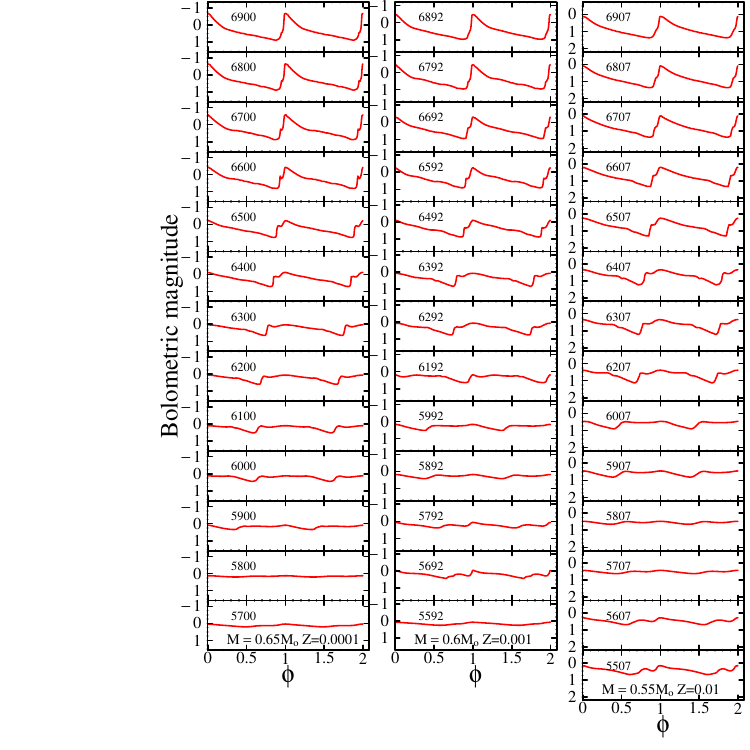}
    \caption{\label{fig:bolcurve} Example of predicted bolometric light curves with decreasing effective temperature and the highest computed stellar mass (see labels) for each metallicity $Z=0.0001$ (left panel), $Z=0.001$ (middle panel) and $Z=0.01$ (right panel).}
\end{figure*}

\begin{figure*}[ht]
    \includegraphics[trim=2.4cm 0.0cm 0cm 0.0cm, clip,width=1.0\linewidth]{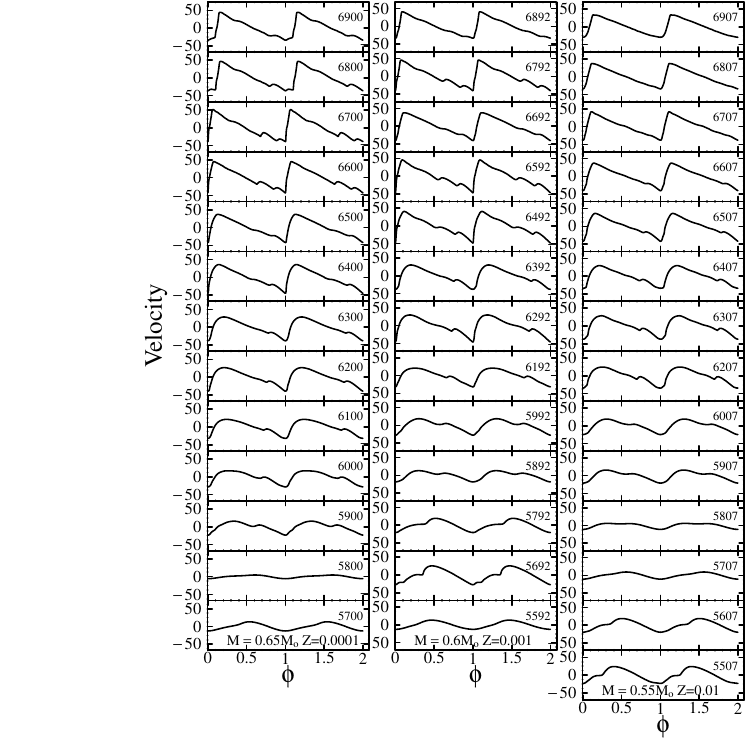}
    \caption{Same as Fig.~\ref{fig:bolcurve} but for radial velocity curves.}\label{fig:velcurve} 
\end{figure*}

In Fig.~\ref{fig:bolcurve} we show an example of the produced atlas of bolometric light curves, whereas
Figure~\ref{fig:velcurve} shows the radial velocity curves for the same models.

\begin{figure*}[ht]
    \includegraphics[trim=2.4cm 2.5cm 0cm 0.0cm, clip,width=0.5\linewidth]{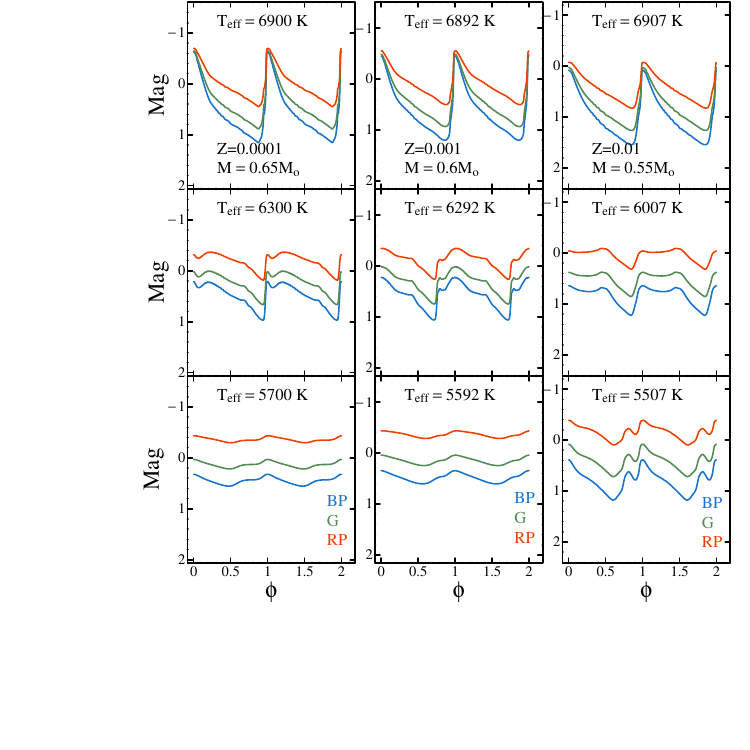}
    \caption{\label{fig:gaiacurve} An example of predicted $\rm G_{BP}$ (red), $\rm G$ (green) and $\rm G_{RP}$ (blue) light curves for $Z=0.0001$ (left panel), $Z=0.001$ (middle panel) and $Z=0.01$ (right panel). The three vertical panels correspond to different values of the model stellar mass and metallicity (see labels). Three values of the effective temperature are considered and labelled in each panel.}
\end{figure*}

\begin{figure*}[ht]
    \includegraphics[trim=2.4cm 2.5cm 0cm 0.0cm, clip, width=0.5\linewidth]{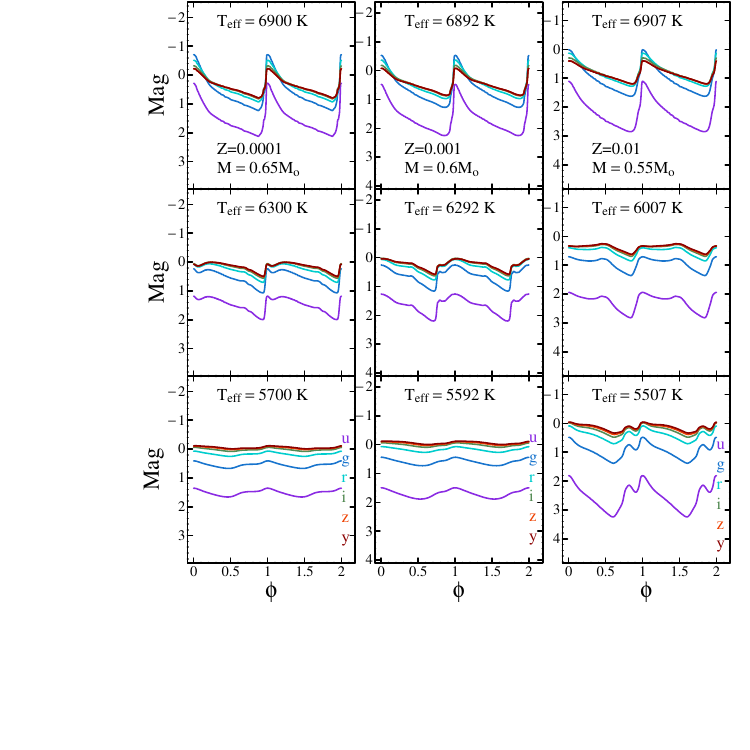}
\caption{\label{fig:lsstcurve} The same as Fig.~\ref{fig:lsstcurve} but for the Rubin-LSST filters.}
\end{figure*}

In Fig.~\ref{fig:gaiacurve} and ~\ref{fig:lsstcurve} we show the same curves already displayed in Fig.~\ref{fig:bolcurve} but in the Gaia and Rubin-LSST bands, respectively.

\begin{figure*}[ht]
    \includegraphics[trim=2.8cm 6.5cm 0cm 0.0cm, clip,width=\textwidth]{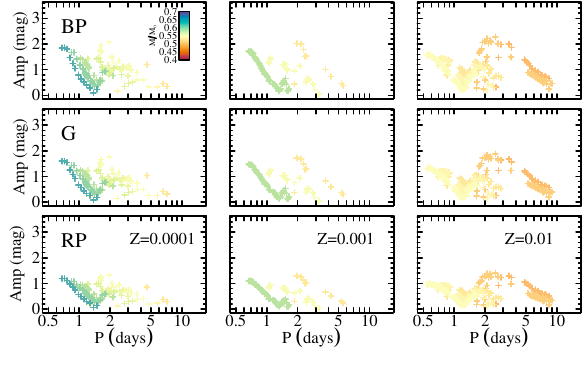}
    \caption{Predicted period-amplitude diagrams for the Gaia bands (top:$G_{BP}$, middle: $G$, bottom: $G_{RP}$) for all the computed models and the labelled metal abundances (left: Z=0.0001, center: Z=0.001, right: Z= 0.01), colour-coded by the assumed stellar mass.}\label{fig:gaiabailey} 
\end{figure*}

\begin{figure*}[ht]
    \includegraphics[width=\textwidth]{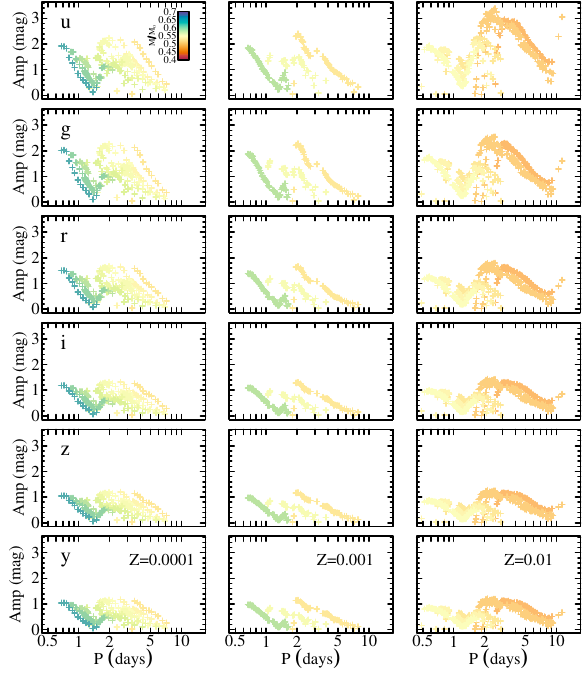}
    \caption{Same as Fig.~\ref{fig:gaiabailey} but for the 6 Rubin-LSST bands labelled in the first column of each row. }\label{fig:lsstbailey}
\end{figure*}

In Fig.~\ref{fig:gaiabailey} we show the predicted period-amplitude diagram in the three Gaia bands for all the computed models and the labelled metal abundances. The same diagram but for the Rubin-LSST bands is shown in Fig.~\ref {fig:lsstbailey}.

\end{appendix}

\end{document}